\documentclass[journal=nalefd,manuscript=letter,layout=traditional]{achemso}

\usepackage[version=3]{mhchem} 
\usepackage[T1]{fontenc}       
\usepackage{graphicx}
\usepackage{amsmath}

\author{Santiago J. Cartamil-Bueno}
\email{s.j.cartamilbueno@tudelft.nl}
\affiliation[Delft University of Technology]
{Kavli Institute of Nanoscience, Delft University of Technology, Lorentzweg 1, 2628CJ, Delft, The Netherlands}
\author{Peter G. Steeneken}
\affiliation[Delft University of Technology]
{Kavli Institute of Nanoscience, Delft University of Technology, Lorentzweg 1, 2628CJ, Delft, The Netherlands}
\author{Alba Centeno}
\affiliation[Graphenea]{Graphenea SA, 20018 Donostia-San Sebasti\'an, Spain}
\author{Amaia Zurutuza}
\affiliation[Graphenea]{Graphenea SA, 20018 Donostia-San Sebasti\'an, Spain}
\author{Herre S. J. van der Zant}
\affiliation[Delft University of Technology]
{Kavli Institute of Nanoscience, Delft University of Technology, Lorentzweg 1, 2628CJ, Delft, The Netherlands}
\author{Samer Houri}
\email{s.houri@tudelft.nl}
\affiliation[Delft University of Technology]
{Kavli Institute of Nanoscience, Delft University of Technology, Lorentzweg 1, 2628CJ, Delft, The Netherlands}

\title[Colorimetry]
  {Colorimetry technique for scalable characterization of suspended graphene}


\begin{document}

\begin{tocentry}
	~~~
	\graphicspath{{Supporting_Information/}}
	\includegraphics{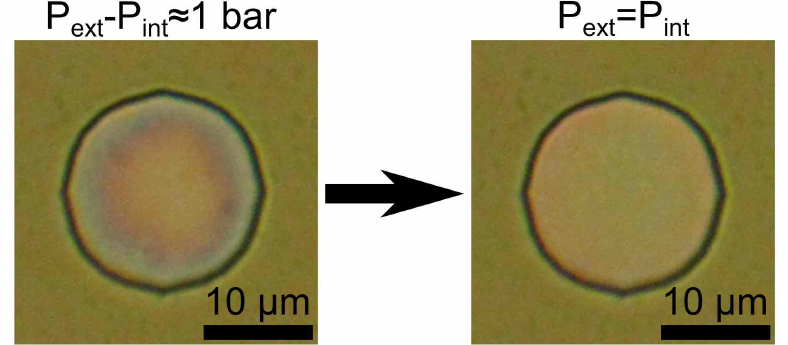}
\end{tocentry}

\begin{abstract}
Previous statistical studies on the mechanical properties of chemical-vapor-deposited (CVD) suspended graphene membranes have been performed by means of measuring individual devices or with techniques that affect the material. Here, we present a colorimetry technique as a parallel, non-invasive, and affordable way of characterizing suspended graphene devices. We exploit Newton's rings interference patterns to study the deformation of a double-layer graphene drum 13.2~$\mu$m in diameter when a pressure step is applied. By studying the time evolution of the deformation, we find that filling the drum cavity with air is 2-5 times slower than when it is purged.
\end{abstract}

\section{Introduction}

Graphene, a monolayer of carbon atoms in honeycomb configuration, has become a subject of active study since its discovery in 2004 \cite{Novoselov2004}. Many potential applications that exploits its electromagnetic properties \cite{Koppens2011} and gas impermeability \cite{Bunch2008} have been proposed, from electronic switching \cite{Yang2012} to water filtration \cite{Nair2012} and pressure sensing \cite{Smith2013,Dolleman2015}. Suspending graphene on circular cavities or trenches eliminates the negative impact of the substrate in its electrical conductivity \cite{Rickhaus2015}, which even allows the graphene to emit visible light \cite{Kim2015}, and enables electro- or opto-mechanical actuation for mass, force and position sensing \cite{Eichler2011,Ferrari2014}. However, graphene suspended movable devices are fragile and tend to break or collapse during the fabrication or measurement \cite{Suk2011,Loez-Polin2015}.

Single-layer graphene (SLG) drums have been extensively studied, showing unique mechanical properties \cite{Frank2007,Meyer2007,Bunch2007,Poot2008,Lee2008a}. Many groups have demonstrated the scalability of CVD SLG drums and analyzed the statistical variation of their mechanical properties by measuring several drums with laser interferometry \cite{Zande2010,Barton2011}, Raman spectroscopy \cite{Suk2011,Metten2014,Shin2016}, and atomic force microscopy \cite{Lee2013,Hwangbo2014}. However, any attempt to commercialize graphene mechanical sensors is ineffective unless a characterization technique that is parallel, contactless, and affordable at the same time becomes available. Furthermore, CVD SLG usually contains gas permeable lattice defects and nanoscale pores due to its growth on imperfect substrates \cite{OHern2012}, which blocks its application in gas pressure sensing devices that require impermeable membranes. A possible route to overcome this difficulty is to stack several CVD layers to reduce the probability of having nanopores from different layers aligned on the same spot \cite{Celebi2014}. 

In this work, we introduce a new non-invasive optical technique to characterize the mechanical properties and the permeability of large arrays of suspended graphene membranes that is similar to Cornu's interference method \cite{Jessop1921}. We observe Newton's rings on a suspended CVD double-layer graphene (DLG) drumhead when applying a pressure difference between inside and outside of the cavity, which allows us to study the deformation of this mechanical system. We find that the rate of volume change is different for purging and filling the cavity with air. Based on these observations, the permeability of the DLG membrane is determined and found to be similar to that of pristine graphene.

Device fabrication starts with a silicon substrate covered with 600~nm of thermally-grown \ce{SiO2}. Circular cavities of different diameters are patterned and etched through the oxide by means of reactive ion etching. Afterwards, a double-layer graphene, made by stacking two SLG CVD graphene layers, is transferred onto the \ce{SiO2}/Si substrate using a semi-dry transfer technique, thus resulting in suspended CVD-DLG drums.
In this paper, we use DLG membranes because its absorption in the visible spectrum of light is at least twice that of SLG, which increases its reflectance while remaining highly transparent \cite{Ochoa-Martinez2015}.

\begin{figure}[ht]
	\graphicspath{{Figures/}}
	\includegraphics{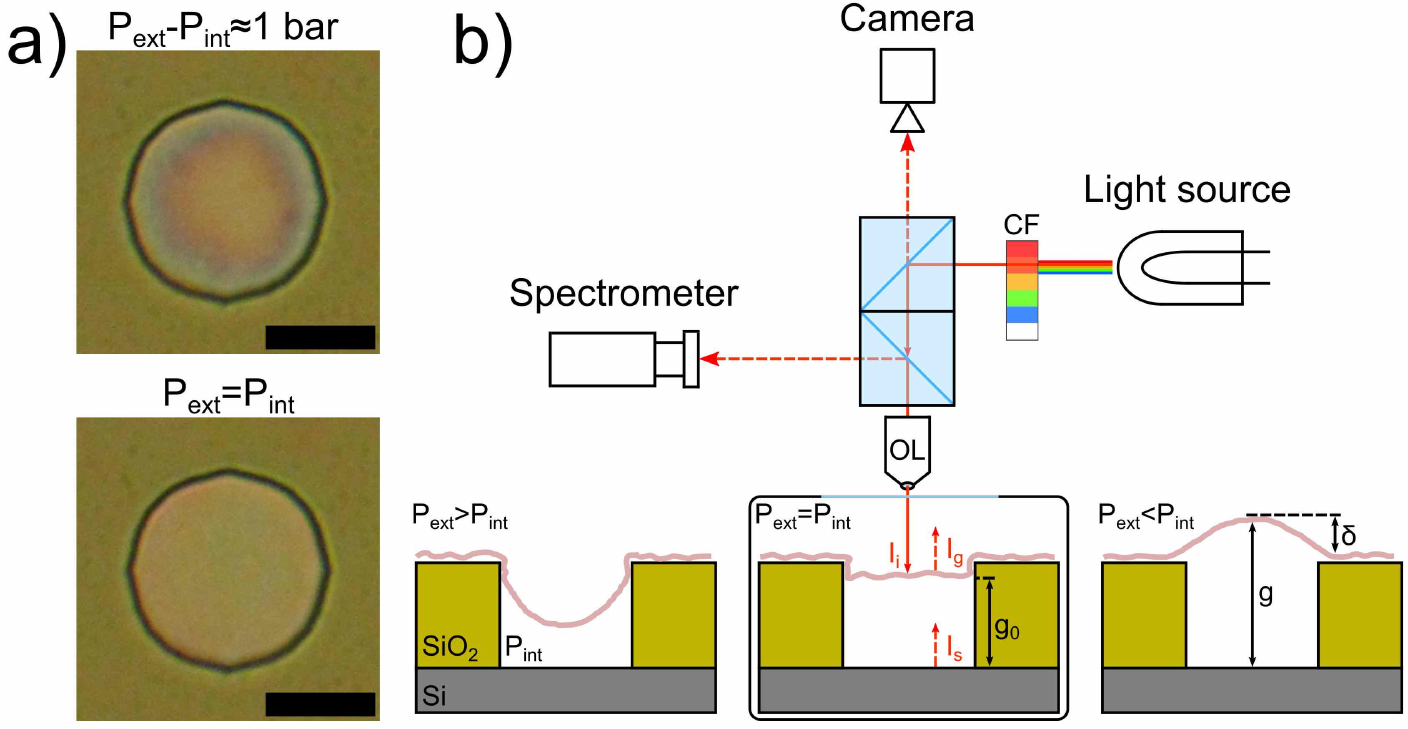}
	\caption{Newton's rings in a graphene drum and colorimetry setup.
		(a) Top: White light camera image of a downward deflected CVD double-layer graphene drum (13.2~$\mu$m in diameter) showing Newton's rings when the pressure difference between outside and inside the cavity is about 1~bar. Pressure-induced deflection of the drumhead changes the gap distance of the optical cavity radially.
		Bottom: After a certain time in an atmospheric environment, the cavity pressure is balanced and the drumhead is flat again, showing an homogeneous color. Scale bars are 10~$\mu$m.
		(b) Colorimetry setup: a customized optical microscope with a 100x objective lens (OL) and K\"{o}hler illumination is used for imaging the suspended graphene drum. White light from a Halogen lamp goes through a color filter (CF), and the intensity reflected from the drum $I_g$ and the silicon substrate $I_s$ is collected by a calibrated consumer camera and a spectrometer. Depending on the pressure difference, the drumhead bends downwards (left panel), remains flat (center panel) or bulges up (right panel).}
	\label{fgr:Fig1}
\end{figure}

Upon the application of a pressure difference between the inside of the cavity $P_{int}$ and the outside of the cavity $P_{ext}$, the circular membrane is deformed. The drum deflects inward (upward) if the pressure difference $\Delta P = P_{ext}-P_{int}$ is positive (negative).
Figure~\ref{fgr:Fig1}a shows an image of a DLG drum of 13.2~$\mu$m in diameter under white light illumination with $\Delta P \approx$1~bar (top) and $\Delta P$=0 (bottom). When the membrane is highly deflected as in Figure~\ref{fgr:Fig1}a, top panel, the circular geometry of the device causes the creation of concentric rings, which are also known as Newton's rings \cite{Georgiou2011}. Progressively with time, air fills the cavity until $P_{ext}$ and $P_{int}$ become equal, resulting in a homogeneous color across the whole drum (Figure~\ref{fgr:Fig1}a, bottom panel).

When illuminated, part of the light is reflected from the suspended membrane $I_g$, and interferes with the light that crossed the membrane and is reflected from the bottom of the cavity $I_{s}$.
This interference is constructive or destructive depending on the illumination wavelength and the distance separating the two surfaces. Since the absorption and internal reflections of the circular DLG membrane play a small role in the interference, the total reflected intensity for a given wavelength $\lambda$ at a distance $r$ from the center would be approximately $ I(r) = I_{g} + I_{s} + 2 \sqrt{I_{g} I_{s}} \cos \left(\frac{4 \pi}{\lambda} \, (g_0-\delta(r)) + \phi \right)$, where $g_0$ is the gap distance between the non-deflected membrane and the bottom of the cavity, $\delta(r)$ is the radial deflection, and $\phi$ is the phase change induced by the reflecting surfaces \cite{Blake2007}.

To fully exploit this phenomenon for fast mechanical and permeability characterization of suspended membranes, we use monochromatic illumination. For this purpose, a cost-effective customized optical microscope setup with K\"{o}hler illumination using a simple white light Halogen lamp (Thorlabs OSL2) as an effective multi-wavelength source is built as shown in Figure~\ref{fgr:Fig1}b. A consumer camera (Canon EOS 700D) collects the total reflected intensity from each point of the drum that is placed inside a vacuum chamber.
Monochrome RAW images are taken at a particular wavelength by using color filters (FWHM 4-10~nm). The intensity values from the CMOS camera are calibrated to correct for gamma compression \cite{Renshaw1991} (for more details, see Supporting Information).
Furthermore, the reflectance is determined by applying $\mathcal{R} = \frac{I}{I_{s}}\mathcal{R}_{s}$, where $\mathcal{R}_{s}$ is the silicon reflectivity \cite{Blake2007}, and $I_{s}$ is the reflected intensity measured from an uncovered circular cavity. This normalization also cancels the inhomogeneous spectral intensity of the source and eliminates the impact of changes in the illumination intensity over time.

\begin{figure}[ht]
	\graphicspath{{Figures/}}
	\includegraphics{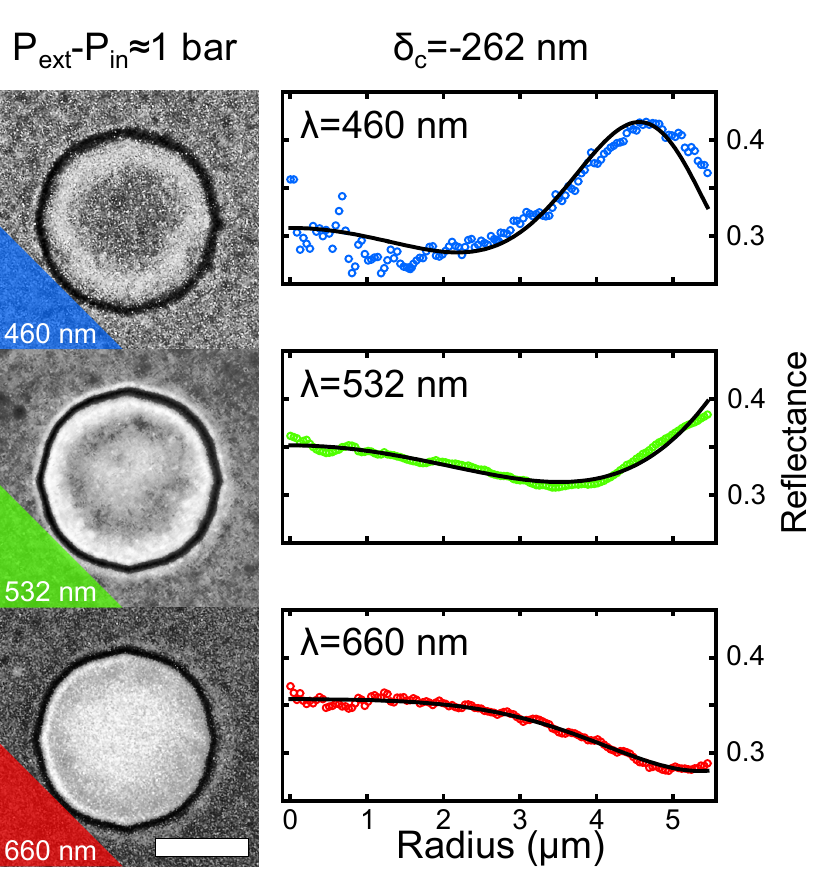}
	\caption{Radial reflectance profile across the drum.
		On the left, reflectance profile of the drum for different wavelengths at the same pressure difference as in Figure~\ref{fgr:Fig1}a (top). Scale bar is 10~$\mu$m.
		On the right, the radial-averaged reflectance (circles) of the corresponding wavelengths. The Newton's rings fit (line) gives a center deflection value of $\delta_c$=-262~nm.}
	\label{fgr:Fig2}
\end{figure}

The radial deflection function of a circular membrane with radius $R$ can be approximated by the maximum deflection at the center $\delta_c$ multiplied by the profile function $f(r) = 1-0.9 \frac{r^2}{R^2} -0.1 \frac{r^5}{R^5}$ \cite{Young2002}. Then, the reflectance as a function of $r$ becomes

\begin{equation}
	\mathcal{R}(r) \propto \cos \left(\frac{4 \pi}{\lambda} \, (g_0-\delta_c \, f(r)) + \phi \right) \\
	= \cos\left({ \Omega \, f(r)} + \phi '\right) \\,
	\label{eqn:Eq1}
\end{equation}

\noindent
where $\Omega = -\frac{4 \pi}{\lambda} \, \delta_c$ and $\phi '= \frac{4 \pi}{\lambda} g_0 + \phi $ are the spatial frequency and phase of the Newton's rings. Therefore, $\mathcal{R}(r)$ depends on the wavelength and the maximum deflection. Figure~\ref{fgr:Fig2}a shows three monochromatic images of the drum for different wavelengths ($\lambda$=460~nm, 532~nm, and 660~nm) at $\Delta P \approx$1~bar. Note that shorter wavelengths result in higher spatial frequency $\Omega$. To improve the data analysis, especially for the short-wavelength noisy images, we take advantage of the centro-symmetry of the device and perform a radial average as shown in the figure for the corresponding monochromatic images.
By fitting the radial-averaged reflectance to Eq.~\ref{eqn:Eq1}, it is possible to obtain a value of $\Omega$ and hence $\delta_c$. In the case of the Newton's rings in Figure~\ref{fgr:Fig2}, the extracted center deflection is $\delta_c$=-262~nm.

\begin{figure}[ht]
	\graphicspath{{Figures/}}
	\includegraphics{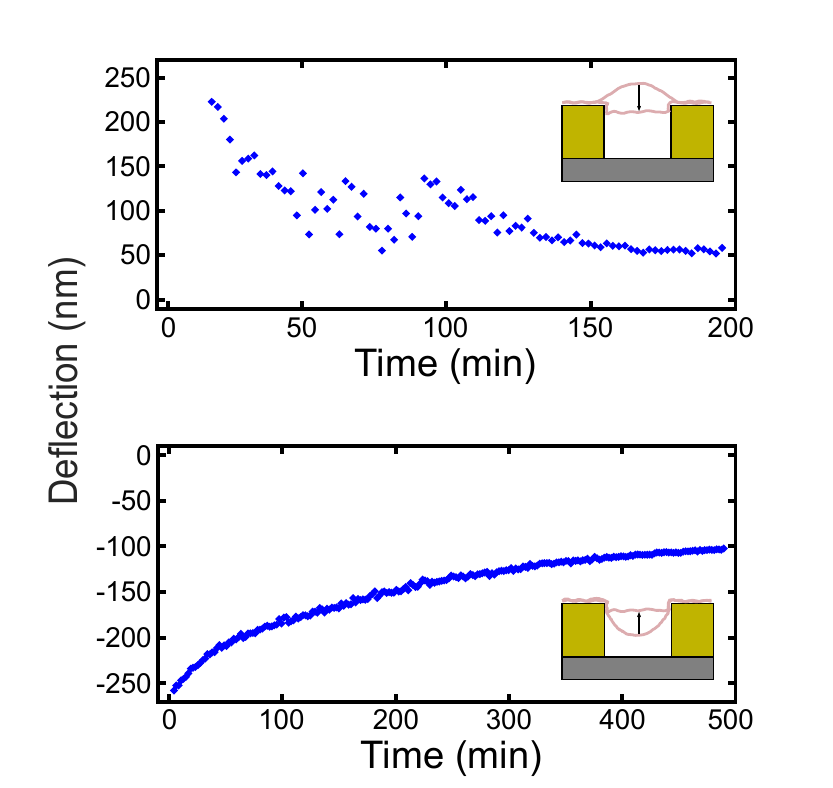}
	\caption{Time evolution of the deflection during purging and filling of the cavity.
		(a) The Newton's rings fitting is applied to a series of reflectance profiles of the same drum for $\lambda$=532~nm, resulting in the change of deflection while the drum returns to pressure balance (blue points). When the gas is escaping the cavity (top), the deflection returns to zero faster than in when the gas is filling the cavity (bottom). The Newton's rings fitting becomes inaccurate at small deflections.}		
	\label{fgr:Fig3}
\end{figure}

The Newton's rings fitting can be used to study the change in pressure-induced deflection with time, and therefore measure the permeability of the drum. The time-dependent maximum deflection as obtained for $\lambda$=532~nm illumination is shown in Figure~\ref{fgr:Fig3} for both purging the air (top) and filling the cavity (bottom). We observe that the gas escapes from the cavity faster than when it has to fill it. As a technical remark, we also observe that the value of $\delta_c$ saturates around 70~nm due to the inaccuracy of fitting the data to Eq.~\ref{eqn:Eq1} in the limit of small deflections.
To understand these rate differences we model the gas filling and purging of the cavity using the ideal gas law, and Hooke's law considering only the non-linear term for large deflections \cite{Koenig2012}. The molecular flux from/to the inside of the cavity at large deflections is

\begin{equation}
\frac{\mathrm{d}n}{\mathrm{d}t} = \frac{ 4 b k_3 \delta ^3 + 3 g_0 k_3 \delta ^2 + A b P_{ext}}{R T} \, \frac{\mathrm{d}\delta}{\mathrm{d}t},
\label{eqn:Eq2}
\end{equation}

\noindent
with positive deflection when purging and negative deflection when filling, and where $R$ is the universal gas constant, $T$=300~K is the temperature, $k_3$ is the nonlinear spring constant of the drum and $A$, its area; $b$=0.52 is a geometric volume factor from the profile of a deflected circular membrane \cite{Koenig2011} (see Supporting Information). Equation~\ref{eqn:Eq2} has two different solutions that depend on whether the drum is bulged upward or inward. Furthermore, the equation predicts that the molecular flux at large deflections depends on the external pressure: when the chamber is pumped ($P_{ext} \approx 0$), the gas escapes the drum cavity faster than when the chamber is vented ($P_{ext} \approx$ 1~bar) and the cavity is slowly filled.
This difference in behavior between purging and filling of the cavity is corroborated in Figure~\ref{fgr:Fig3} qualitatively. 

\begin{figure}[ht]
	\graphicspath{{Figures/}}
	\includegraphics{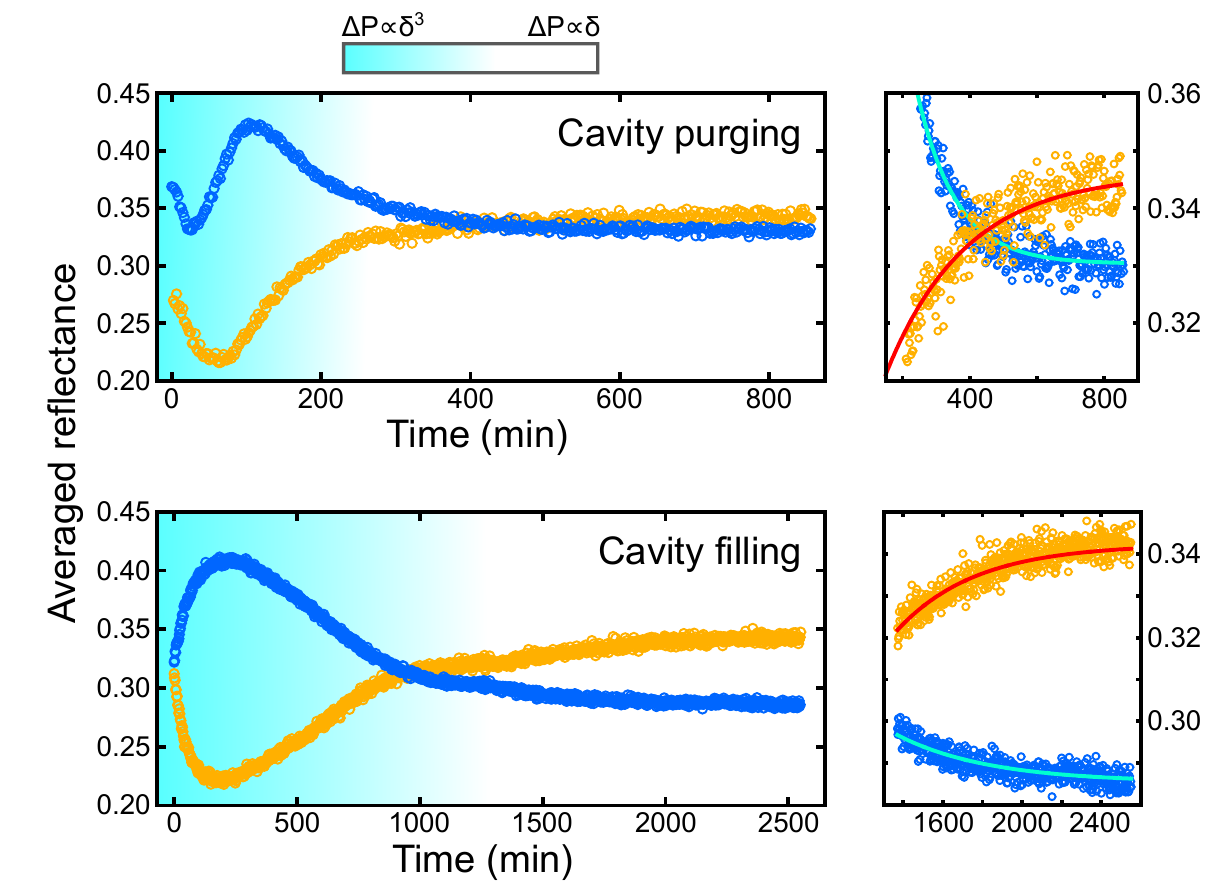}
	\caption{Change of drum averaged reflectance with time during purging and filling: time constant extraction.
	A simple exponential function is fitted to the linear-deflection regime of the drum averaged reflectance (blue circles for $\lambda$=460~nm and orange circles for $\lambda$=600~nm) to obtain the time constants of the drum when purging and filling (values showed in Table~\ref{tbl:table_tau}). The extracted time constant ranges from 125 to 573 minutes, leading to a membrane permeability of around $5 \times 10^{-24} \frac{mol}{s \cdot Pa}$.}
	\label{fgr:Fig4}
\end{figure}

To extract the permeability quantitatively, we consider the small deflection limit. In this linear-deflection regime, the cavity volume can be assumed to be constant and the mechanical response of the drum is governed by its linear stiffness. Under these conditions the membrane deflection is given by a simple decaying exponential $ \delta (t) = \delta_\infty + \delta_0 e^{-t/\tau}$, where ${\tau = \frac{A g_0}{\mathcal{P} R T}}$ is the time constant of the system, $\mathcal{P}$ is the permeability of the membrane, and $\delta_\infty$, its position at rest which does not need to be zero due to side-wall adhesion \cite{Liu2014,Suk2011}.
In addition, for small deflections, the drum reflectance averaged over its area $\mathcal{R}_{drum}$ is directly proportional to the deflection, and can therefore be expressed as $\mathcal{R}_{drum}(t) = A + B \, e^{-t/\tau}$ (see Supporting Information for the derivation).
The drum averaged reflectance is shown in Figure~\ref{fgr:Fig4} for two wavelengths, 460~nm (blue circles) and 600~nm (orange circles), for both cavity filling and purging. As the large deformation regime subsides after 200 minutes for the purging experiment and 1000 minutes for the filling experiment, the reflectance in the linear regime decays exponentially with time. This part of the curves is then fitted to obtain the time constant of the drum. The fitted time constants and the inferred permeability values are summarized in Table~\ref{tbl:table_tau} for different wavelengths for the two experiments.

\begin{table}
	\caption{Values of time constants and permeabilities.}
	\label{tbl:table_tau}
	\begin{tabular}{cc|cccc|cccc}
		\multicolumn{2}{c|}{} & \multicolumn{4}{c|}{Filling} & \multicolumn{4}{c}{Purging} \\
		\hline		
		\multicolumn{2}{c|}{Wavelength (nm)}  & \multicolumn{2}{c}{$\tau$ (min)} & \multicolumn{2}{c|}{$\mathcal{P}$ ($10^{-24} \frac{mol}{s \cdot Pa}$)} & \multicolumn{2}{c}{$\tau$ (min)} & \multicolumn{2}{c}{$\mathcal{P}$ ($10^{-24} \frac{mol}{s \cdot Pa}$)}  \\
		\hline
		\multicolumn{2}{c|}{460}  & \multicolumn{2}{c}{446} & \multicolumn{2}{c|}{4.92} & \multicolumn{2}{c}{125}  & \multicolumn{2}{c}{17.56} \\
		\multicolumn{2}{c|}{532}  & \multicolumn{2}{c}{653} & \multicolumn{2}{c|}{3.36} & \multicolumn{2}{c}{125}  & \multicolumn{2}{c}{17.56} \\
		\multicolumn{2}{c|}{600}  & \multicolumn{2}{c}{519} & \multicolumn{2}{c|}{4.23} & \multicolumn{2}{c}{237}  & \multicolumn{2}{c}{9.26} \\
		\multicolumn{2}{c|}{633}  & \multicolumn{2}{c}{573} & \multicolumn{2}{c|}{3.83} & \multicolumn{2}{c}{6294} & \multicolumn{2}{c}{0.35} \\
		\multicolumn{2}{c|}{660}  & \multicolumn{2}{c}{570} & \multicolumn{2}{c|}{3.85} & \multicolumn{2}{c}{565}  & \multicolumn{2}{c}{3.88} \\
		\hline
	\end{tabular}
\end{table}


Air diffusion must occur through the graphene material because it is unlikely that it leaks through the interface between the DLG membrane and the \ce{SiO2} substrate as the CVD DLG covers large areas. The permeability for the DLG membrane is calculated to be $(4.04 \pm  0.58) 10^{-24} \frac{mol}{s \cdot Pa}$ from the filling data, and $(9.72 \pm  7.83) 10^{-24} \frac{mol}{s \cdot Pa}$ from the purging data, which are one order of magnitude lower than the permeability measured in pristine SLG and BLG \cite{Koenig2012}.
The spread in permeability values in Table~\ref{tbl:table_tau} is larger in the purging experiment than in the filling case due to illumination difficulties that happen during the purging experiment and inaccuracies in the exponential fitting as seen in Table~\ref{tbl:table_tau} that can be large for certain wavelengths.
Nevertheless, the difference between purging and filling the cavity arising from the difference of molecular flux at large deflections is reproducible, which bring us to the conclusion that the change in volume of these ultrathin microdrums cannot be ignored. The effect becomes strong when the membranes are deflected beyond the linear regime, which is usually the case when performing AFM nanoindentation or pressure studies.
To extend these conclusions to other drums and study the dependence with their diameter, we can apply the colorimetry technique to hundreds of drums (see Supporting Information).

This study has presented the colorimetry technique as a scalable tool to characterize the mechanical properties and the permeability of suspended graphene microdevices. This non-invasive optical technique allows to extract the evolution of a thin membrane deflection with time when filling and emptying the microcavity with air. We observe the linear- and nonlinear-deflection regimes, and we find that the gas filling process is slower than the gas purging in the case of large deflection, where volume change is not negligible. Furthermore, we use small deflection measurements to extract the permeability of a double-layer CVD graphene membrane, finding that it is similar to that of pristine SLG and BLG, paving the way to the fabrication of impermeable pressure sensors.
Ultimately, the colorimetry technique could be employed to characterize the mechanical properties of suspended SLG and other 2D materials by using the right combination of cavity depth and wavelength. The parallelization of data acquisition by image processing tools combined with a controlled deformation of the membranes would allow the fast characterization of large arrays of these mechanical systems at once limited only by the resolution of the optical instrument, and could lead to the realization of interferometric modulator displays (IMOD) made out of graphene.

\begin{acknowledgement}
The authors thank W.J. Venstra for useful discussions. The research leading to these results has received funding from the European Union's Horizon 2020 research and innovation programme under grant agreement No 649953 (Graphene Flagship).
\end{acknowledgement}

\begin{suppinfo}

Description of the CVD-SLG growth and CVD-DLG transfer. Gamma factor characterization and correction. Calculation of the geometric volume factor. Newton's rings fitting.
Derivation of the drum averaged reflectance for small deflections. Drum averaged reflectance with time for all the wavelengths for gas purging and filling. Video of the change in color of the drum in a filling experiment. Animated extraction of the reflectance profile as a function of time for a wavelength (660~nm) when filling. Video of the change in color of hundreds of drums in a filling experiment.


\end{suppinfo}

\bibliography{Bibliography}

\newpage

\section{Supporting Information}
\setcounter{figure}{0}
\setcounter{equation}{0}
\renewcommand\thefigure{S-\arabic{figure}}
\renewcommand\theequation{S-\arabic{equation}}

Supporting Information Outline:

1. Description of the CVD-SLG growth and CVD-DLG transfer.

2. Gamma factor characterization and correction.

3. Calculation of the geometric volume factor.

4. Newton's rings fitting.

5. Derivation of the drum averaged reflectance for small deflections.

6. Drum averaged reflectance with time for all the wavelengths for gas purging and filling.

\subsection{1. Description of the CVD-SLG growth and CVD-DLG transfer.}

Single-layer graphene was grown by  chemical vapour deposition (CVD) using a 4'' cold wall reactor (Aixtron BM). Copper foil was used as the catalyst and a surface pre-treatment was carried out in order to remove the native copper oxide and other impurities. The synthesis was carried out at $1000^{\circ}$C using methane as the carbon source. After the synthesis, the single-layer graphene was coated with a polymer layer and stacked onto a second single-layer graphene by using a semi-dry transfer process. The stacked double-layer CVD graphene was transferred onto 5$\times$5~mm$^2$ \ce{SiO2}/Si substrates containing circular cavities of 1-20~$\mu$m in size and 600~nm in depth by following a semi-dry transfer procedure. Finally, the supporting polymer layer was removed by annealing at $450^{\circ}$C for 2~hours in N$_2$ atmosphere.

\subsection{2. Gamma factor characterization and correction.}

\begin{figure}[ht]
	\graphicspath{{Supporting_Information/}}
	\includegraphics{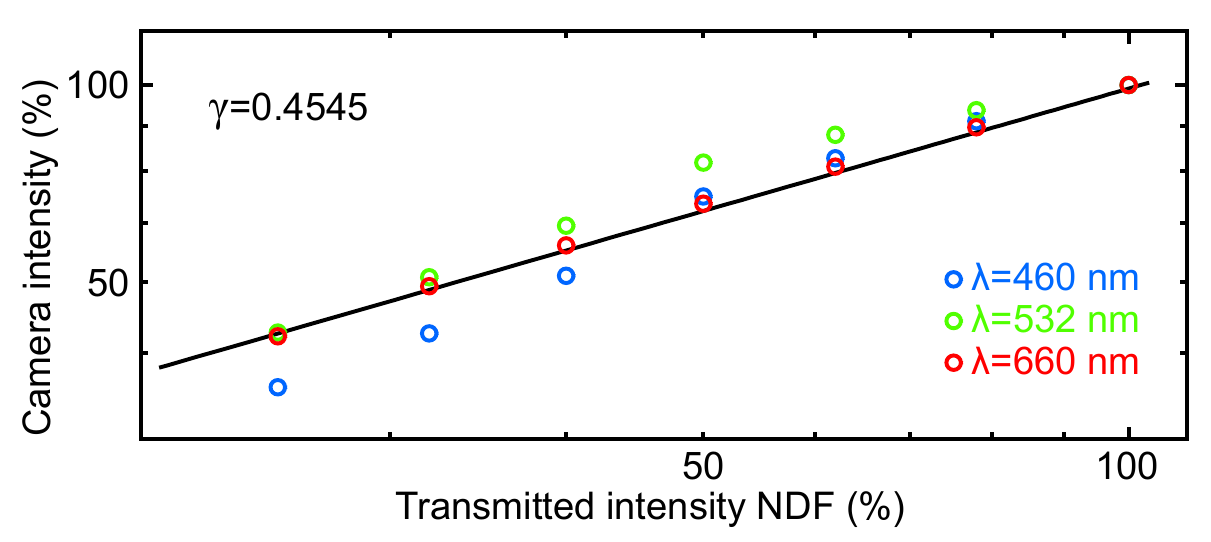}
	\caption{Gamma characterization.}
	\label{fgr:FigS2}
\end{figure}

Consumer CMOS cameras apply an artificial compression to the received intensity values. To characterize the gamma factor, we measured the intensity values reflected from the \ce{SiO2} substrate for different illumination intensities by using neutral density filters. Figure~\ref{fgr:FigS2} shows the power relation of the curves for three wavelengths. We obtain an average gamma of 0.4545 for all the wavelengths.

In the main text, we compensate for the compression by applying a gamma correction
\begin{equation}
I = 255 \left (\frac{I_{RAW}}{255} \right)^{2.2}.
\label{eqn:EqS1}
\end{equation}

\subsection{3. Calculation of the geometric volume factor.}
The volume of a membrane with a radial profile described by the membrane function $f(r) = 1-0.9 \frac{r^2}{R^2} -0.1 \frac{r^5}{R^5}$ is obtained by integration. Thus,

\begin{equation}
\begin{split}
V &= \delta_c \int_0^{2 \pi} \int_0^R  \left(1-0.9 \frac{r^2}{R^2} -0.1 \frac{r^5}{R^5}\right) \, r \mathrm{d}r \mathrm{d}\theta \\
&= \pi R^2 \, 0.52 \, \delta_c.
\end{split}
\end{equation}

Therefore, $V = A \, b$, where $A=\pi R^2$ and $b=0.52$.

\subsection{4. Newton's rings fitting.}
Figure~\ref{fgr:FigS4} shows the Newton's rings reflectance data fitted to Eq.~\ref{eqn:Eq1} for different deflections for both purging the air and filling the cavity ($\lambda$=532~nm). The Newton's rings fitting is used in the main text to study the change in pressure-induced deflection with time, and therefore measure the permeability of the drum (Figure~\ref{fgr:Fig3}).

\begin{figure}[ht]
	\graphicspath{{Supporting_Information/}}
	\includegraphics{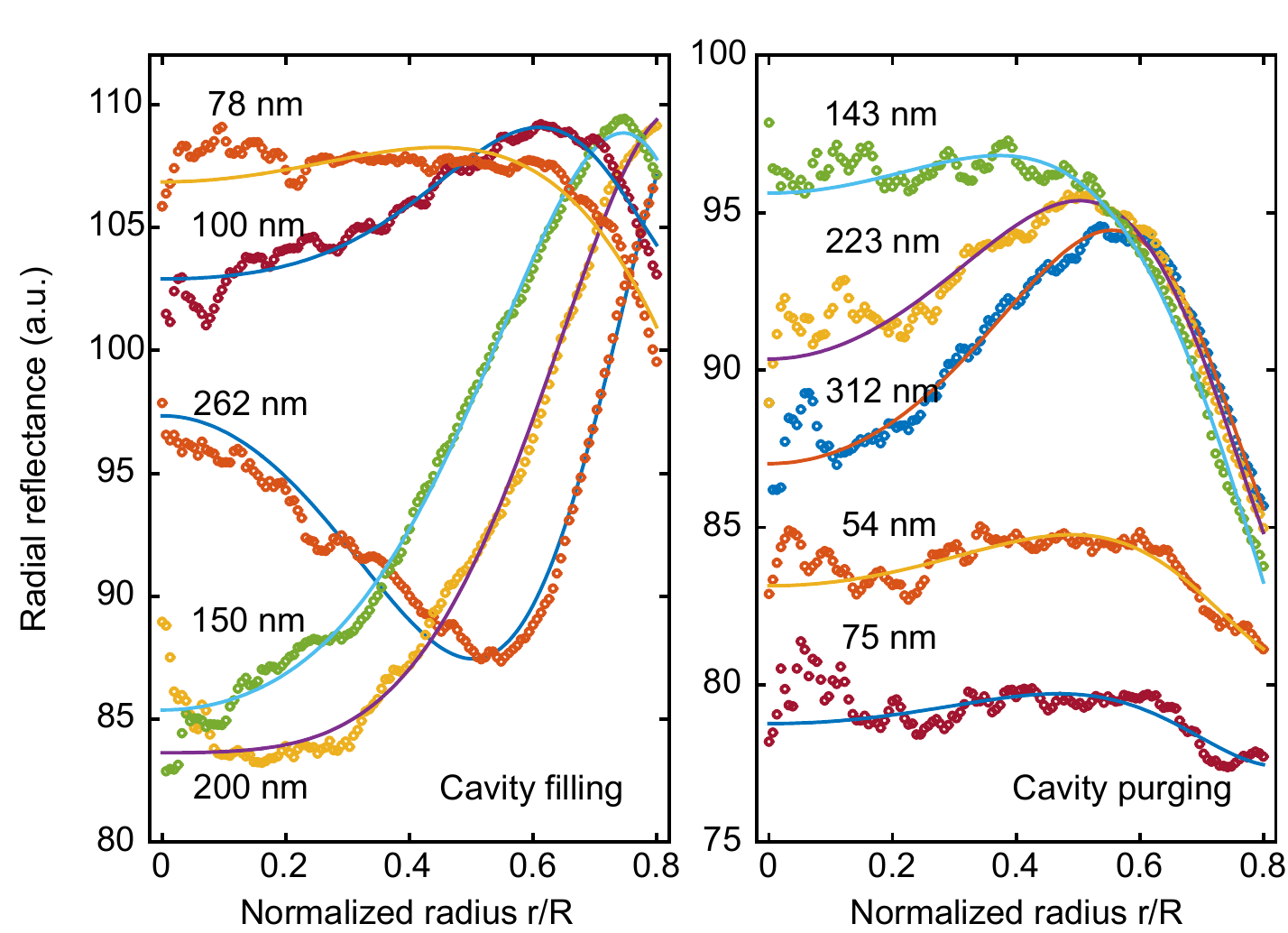}
	\caption{Newton's rings fitting for different deflections.}
	\label{fgr:FigS4}
\end{figure}

\subsection{5. Derivation of the drum averaged reflectance for small deflections.}

The radial reflectance is
\begin{equation}
\mathcal{R}(r,t) \approx R_s + R_g + 2 \sqrt{R_g R_s} \cos\left(\frac{4 \pi}{\lambda} \, (g_0-\delta(r,t)) + \phi \right) = R_s + R_g + 2 \sqrt{R_g R_s} \cos({ \Omega(t) \, f(r)} + \phi ')  \\,
\end{equation}

\noindent
where $\Omega = -\frac{4 \pi}{\lambda} \, \delta_c(t)$ and $\phi '= \frac{4 \pi}{\lambda} g_0 + \phi $ are the spatial frequency and phase of a Newton ring. $f(r) = 1-0.9 \frac{r^2}{R^2} -0.1 \frac{r^5}{R^5}$ is a profile function that can be approximated to $f(r) \approx 1-\frac{r^2}{R^2}$ for small deflections. Then,
\begin{equation}
\mathcal{R}(r,t) \approx a + b \, \cos\left(-{ \frac{4 \pi}{\lambda} \, \delta_c(t) \, \left(1-\frac{r^2}{R^2}\right)} + \phi '\right) \\.
\end{equation}

Taking the surface integral and using $r \mathrm{d}r = \frac{R^2}{2} \mathrm{d}\left(\frac{r^2}{R^2} -1\right)=\frac{R^2}{2} \mathrm{d}X$
\begin{equation}
\begin{split}
\int_{S} \mathcal{R}(r,t) \, ds &= \int_0^{2 \pi} \int_0^R  \mathcal{R}(r,t) \, r \mathrm{d}r \mathrm{d}\theta \\
&\approx 2 \pi \int_0^R  \left[a + b \, \cos\left({ \frac{4 \pi}{\lambda} \, \delta_c(t) \, \left(\frac{r^2}{R^2} -1\right)} + \phi '\right)\right] \, r \mathrm{d}r \\
&\approx \pi R^2 a + \pi R^2 b \int_{-1}^0  \cos\left({ \frac{4 \pi}{\lambda} \, \delta_c(t) \, X} + \phi '\right) \, \mathrm{d}X \\
&\approx \pi R^2 a + \pi R^2 b \frac{\lambda}{4 \pi \delta_c(t)} \left[\sin(\phi ') - \sin\left(\phi ' - \frac{4 \pi}{\lambda} \, \delta_c(t)\right)\right]. \\
\end{split}
\end{equation}

We can expand the second sine
\begin{equation}
\sin\left(\phi ' - \frac{4 \pi}{\lambda} \, \delta_c(t)\right) \approx \sin(\phi ') - \frac{4 \pi}{\lambda} \, \delta_c(t) \cos (\phi ') - \frac{1}{2} {\frac{4 \pi}{\lambda}}^2 \, \delta_c^2(t)\sin(\phi ').
\end{equation}

Therefore,
\begin{equation}
\begin{split}
\mathcal{R}_{drum}(t) &= a + b \, \frac{\lambda}{4 \pi \delta_c(t)} \left(\sin(\phi ') - \sin(\phi ' - \frac{4 \pi}{\lambda} \, \delta_c(t))\right) \\
&\approx a + b \frac{\lambda}{4 \pi \delta_c(t)} \left(\frac{4 \pi}{\lambda} \, \delta_c(t) \cos (\phi ') + \frac{1}{2} {\frac{4 \pi}{\lambda}}^2 \, \delta_c^2(t)\sin(\phi ')\right) \\
&\approx a + b \, \cos (\phi ') + \frac{2 \pi b}{\lambda} \, \sin(\phi ')\delta_c(t).\\
\end{split}
\end{equation}

For small deflections, we know that the change of deflection with pressure is a decaying exponential $\delta_c(t) = \delta_0 \, e^{-t/\tau}$. Then,
\begin{equation}
\mathcal{R}_{drum}(t) \approx A + B \, e^{-t/\tau},
\end{equation}

\noindent
where $A = a + b \cos (\phi ')$, $B = \frac{2 \pi b}{\lambda} \, \sin(\phi ')\delta_0$, and $\delta_0$ is the initial deflection in the linear regime.

\subsection{6. Drum averaged reflectance with time for all the wavelengths for gas purging and filling.}
Figure~\ref{fgr:FigS4} shows the drum average reflectance data for all the wavelengths for the purging (top) and filling (bottom) cases. Figure~\ref{fgr:Fig4} in the main text uses this data for two wavelengths (460~nm and 600~nm).

\begin{figure}[ht]
	\graphicspath{{Supporting_Information/}}
	\includegraphics{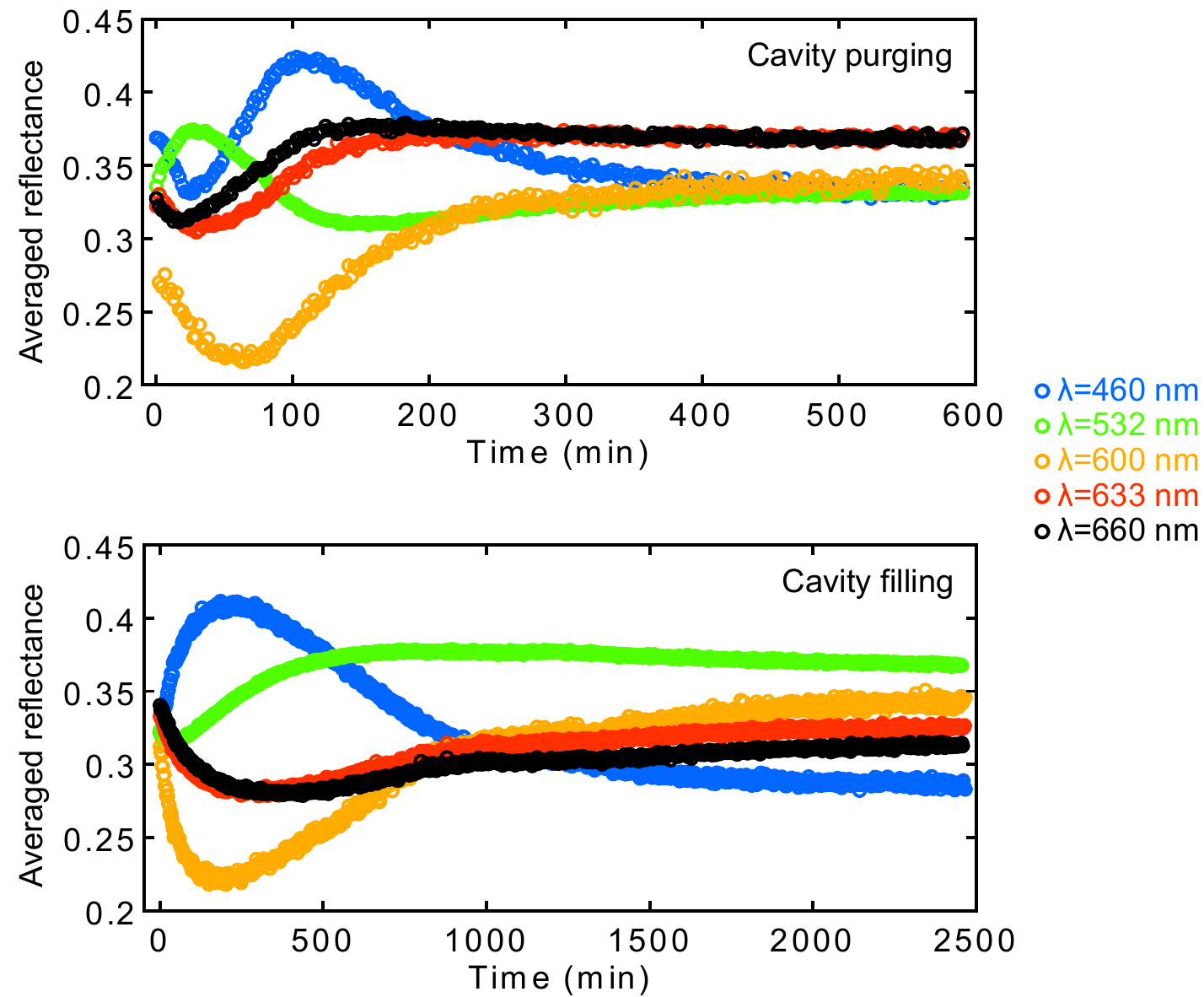}
	\caption{Drum averaged reflectance for all wavelengths.}
	\label{fgr:FigS4}
\end{figure}

\end{document}